\documentclass[a4paper,12pt]{article}
\usepackage{epsfig,amssymb,amsmath}
\usepackage{graphicx}
\usepackage{amsfonts}

\begin{document}

%

\newcommand{\be}{\begin{equation}}
\newcommand{\ee}{\end{equation}}
\newcommand{\bea}{\begin{eqnarray}}
\newcommand{\eea}{\end{eqnarray}}
\newcommand{\bean}{\begin{eqnarray*}}
\newcommand{\eean}{\end{eqnarray*}}
\font\upright=cmu10 scaled\magstep1
\font\sans=cmss12
\newcommand{\ssf}{\sans}
\newcommand{\stroke}{\vrule height8pt width0.4pt depth-0.1pt}
\newcommand{\Z}{\hbox{\upright\rlap{\ssf Z}\kern 2.7pt {\ssf Z}}}
\newcommand{\ZZ}{\Z\hskip -10pt \Z_2}
\newcommand{\C}{{\rlap{\upright\rlap{C}\kern 3.8pt\stroke}\phantom{C}}}
\newcommand{\R}{\hbox{\upright\rlap{I}\kern 1.7pt R}}
\newcommand{\HH}{\hbox{\upright\rlap{I}\kern 1.7pt H}}
\newcommand{\CP}{\hbox{\C{\upright\rlap{I}\kern 1.5pt P}}}
\newcommand{\identity}{{\upright\rlap{1}\kern 2.0pt 1}}
\newcommand{\half}{\frac{1}{2}}
\newcommand{\quart}{\frac{1}{4}}
\newcommand{\pr}{\partial}
\newcommand{\bm}{\boldmath}
\newcommand{\I}{{\cal I}} 
\newcommand{\M}{{\cal M}}
\newcommand{\N}{{\cal N}}
\newcommand{\e}{\varepsilon}

\thispagestyle{empty}
\rightline{DAMTP-2018-39}
\vskip 3em
\begin{center}
{{\bf \Large Forces between Kinks and Antikinks with Long-range Tails
}} 
\\[15mm]

{\bf \large N.~S. Manton\footnote{email: N.S.Manton@damtp.cam.ac.uk}} \\[20pt]

\vskip 1em
{\it 
Department of Applied Mathematics and Theoretical Physics,\\
University of Cambridge, \\
Wilberforce Road, Cambridge CB3 0WA, U.K.}
\vspace{12mm}

\abstract
{In a scalar field theory with a symmetric octic potential having a
quartic minimum and two quadratic minima, kink solutions have 
long-range tails. We calculate the force between two 
kinks and between a kink and an antikink when their long-range 
tails overlap. This is a nonlinear problem, solved using an adiabatic 
ansatz for the accelerating kinks that leads to a modified, 
first-order Bogomolny equation. We find that the kink-kink force 
is repulsive and decays with the fourth power of the kink
separation. The kink-antikink force is attractive and decays similarly.
Remarkably, the kink-kink repulsion has four times the
strength of the kink-antikink attraction.}

\end{center}

\vskip 150pt
\leftline{Keywords: Kinks, Long-range tail, Scalar field, Force}
\vskip 1em

\vfill
\newpage
\setcounter{page}{1}
\renewcommand{\thefootnote}{\arabic{footnote}}


\section{Introduction} 
\vspace{4mm}

Consider a Lorentz-invariant scalar field theory in $1+1$ dimensions
with one real field $\phi(x,t)$ and Lagrangian density
\be
{\cal L} = \half \left(\frac{\pr\phi}{\pr t}\right)^2 
- \half\left(\frac{\pr\phi}{\pr x}\right)^2 - V(\phi) \,.
\label{Lagran}
\ee
Its field equation is
\be
\frac{\pr^2\phi}{\pr t^2} - \frac{\pr^2\phi}{\pr x^2} +
\frac{dV}{d\phi} = 0 \,.
\label{fieldeq}
\ee
Assume that the potential $V(\phi)$ is smooth and 
non-negative, and attains its global
minimum value $V=0$ for one or more values of $\phi$. These field
values are vacua of the theory, and if $V$ has more than one vacuum, 
then it has static kink solutions interpolating between them as $x$ 
increases from $-\infty$ to $\infty$. Much is known about kinks for 
generic $V$ of this type, and also in particular examples \cite{Raj,book,Shn}. 
The classic kink occurs for the quartic potential $V(\phi) = 
\half(1 - \phi^2)^2$. The solution $\phi(x) = \tanh(x)$ 
connects the vacua $\phi = -1$ and $\phi = 1$, and there is an
antikink going the other way. If the theory is 
extended to higher spatial dimensions, then the kink becomes a domain 
wall, but we will not consider this generalisation further.

If $V$ has a finite or infinite sequence
of vacua $\phi_n$ (in increasing order) then there is a
kink connecting $\phi_n$ to $\phi_{n+1}$ and an antikink connecting 
$\phi_{n+1}$ to $\phi_n$ for each $n$, but there are no kinks
connecting vacua whose indices differ by more than one \cite{Raj}. Field 
configurations connecting non-adjacent vacua can be interpreted 
as nonlinear superpositions of more than one kink, but these
configurations are never static. Physically, the kinks repel
each other and separate. Field configurations connecting $\phi_n$ to
close to $\phi_{n+1}$ and then back to $\phi_n$ can be regarded as 
kink-antikink pairs. These are not static either, as a
kink and antikink attract each other. Note that the assumption 
that the kinks and antikinks are interpolating between
vacua is important. It is possible for a kink and antikink to be in
static equilibrium at a finite separation if the asymptotic field
value $\phi$ is metastable, i.e. a local but not global minimum of $V$
\cite{BM}. 

The field equation for a static kink is a second-order ODE, but it can
be reduced to a first-order equation. This is a special case
of the Bogomolny trick \cite{Bo} (although known much earlier in this 
context). We shall refer to the first-order equation as the 
Bogomolny equation for the kink, and use it repeatedly, even 
in the context of kink dynamics.

The minima of a generic potential are all quadratic, 
with $V$ having positive second derivative. In this case
the tails of kinks are short-ranged, i.e. the field $\phi$ approaches
the vacuum values, between which it is interpolating, exponentially 
fast as $x \to \pm\infty$. However, Lohe \cite{Loh}, and more recently 
Khare et al. \cite{KCS} and Bazeia et al. \cite{BMM}, have
drawn attention to several examples where at least one minimum is not
quadratic. This is not generic, but easily occurs as parameters in $V$
are varied. If one of the global minima of $V$ is quartic (the next 
simplest case) then one tail of the kink that approaches it is 
long-ranged; $\phi$ approaches the quartic minimum with a 
$\frac{1}{x}$ behaviour.

When the tails are short-ranged, then it is possible to calculate the
force between two well-separated kinks \cite{book}. 
This is for the situation where the kink on the left (smaller $x$) 
interpolates between $\phi_{n-1}$ and $\phi_n$, and the kink on the 
right interpolates between $\phi_n$ and $\phi_{n+1}$. The calculation
relies on a linear superposition of the exponentially small tails in
the region between the kinks. The result is a force that
decays exponentially fast with the kink separation. An example is 
discussed in Appendix B.

There are a number of ways of approaching the force calculation. 
The force on one of the kinks can be found using a version of 
Noether's theorem to determine the rate of
change of its momentum \cite{Ma5}. This is equivalent to using the
energy-momentum tensor to find the stress exerted on the half-line
containing the kink. An alternative is to attempt to
approximately solve the full, time-dependent field equation. One can
make an ansatz describing an accelerating kink, and then match the
tail of this to the tail of the other
kink \cite{Man}. This determines the acceleration. A cruder
approach is simply to set up a static field configuration that
incorporates both kinks, satisfying the appropriate boundary conditions,
and calculate the energy as a function of the separation. The
negative of the derivative of this energy with respect to the separation 
is an estimate of the kink-kink force. Such a static field
configuration does not of course satisfy the full, time-dependent
field equation (\ref{fieldeq}).

None of these methods is completely straightforward to implement, as
each depends on an ansatz for an interpolating field. The methods might not
agree. One also needs to know where the centres of the kinks are, to have a
precise notion of separation. The centres have to be carefully defined,
especially for kinks with an asymmetric profile, and we shall clarify
in Appendix A what a good definition is. Because of the extended 
character of kinks, any formula for the force usually makes sense 
only to leading order in the separation, even when the separation is
large, and subleading terms are meaningless.

The force between two kinks with long-range, $\frac{1}{x}$ tails has apparently 
not been accurately determined. Certainly, the calculation in \cite{book}
breaks down in this case. The force between a kink and antikink with
these tails was estimated to decay with the fourth power of the separation by
Gonz\'alez and Estrada-Sarlabous \cite{GE,MGGL}. The main purpose of
this paper is to establish a similar result for the kink-kink case,
and to find the numerical coefficients. We shall apply the different 
methods outlined above, and show to what extent they give consistent 
results. Our results show unambiguously that the force decays with the 
fourth power of the kink separation.  

Using the calculated force, we can find an effective equation of
motion for the positions of the kinks. Two kinks repel, so they can approach
slowly from infinity, instantaneously stop at a large 
separation, and move out again to infinity. The separation remains large
throughout, so the effective equation of motion should be reliable. 
The dynamics of a kink and antikink is more complicated; they attract and can
then annihilate, resonate as an oscillating pair, or scatter. 

To test these pictures, it is helpful to perform numerical 
simulations of the kink-kink and kink-antikink dynamics. Recently, 
Belendryasova and Gani \cite{BG} studied kink-antikink 
dynamics numerically in the same model as ours, but their initial configuration
was simply the sum of the kink and antikink fields. In this configuration,
the long-range tail of the kink extends right across the antikink, 
producing an asymptotic field whose difference from the local vacuum value
has the wrong sign. This introduces some unwanted initial energy that converts 
quickly into radiation, and substantially influences the magnitude and
even the sign of the force, as has been clarified by 
Christov et al. \cite{Chr}. Further refinement of
the numerical algorithms is therefore desirable.    

\vspace{5mm}

\section{A Model for a Kink with a Long-range Tail}
\vspace{4mm}

There are an unlimited number of scalar field theories with kinks having
long-range tails. Several of them, arising from a variety of polynomial
potentials $V$, are discussed in refs.\cite{Loh,KCS,BMM} and 
elsewhere. However, in many cases, the kink solution is 
not explicit, so the algebra needed to
investigate the kink's properties is complicated. We shall therefore
focus on the simplest symmetric potential that admits two
related kinks with long-range tails, and their antikinks, and 
calculate the forces between these. Our results should generalise.

The potential we consider is the octic polynomial
\be
V(\phi) = \half (1 - \phi^2)^2 \phi^4 \,.
\ee
This has quadratic minima at $\phi = \pm 1$, and a quartic minimum at
$\phi = 0$. The kink interpolates between $\phi = 0$ and $\phi =
1$, and there is a mirror kink that interpolates between $\phi = -1$
and $\phi = 0$, with the same energy. The antikink interpolates
between $\phi = 1$ and $\phi = 0$, and there is an antimirror-kink too.

The potential $V$ can be expressed in terms of a superpotential $W$ as 
\be
V = \half \left(\frac{dW}{d\phi}\right)^2
\ee
where
\be
\frac{dW}{d\phi} = (1 - \phi^2)\phi^2 \,,
\ee
so
\be
W(\phi) = \frac{1}{3}\phi^3 - \frac{1}{5}\phi^5 + {\rm const.}
\ee
The constant has no significance, but it will be convenient to
set it to $-\frac{2}{15}$, so that $W(1) = 0$ and $W(0) = -\frac{2}{15}$.  

Starting with the Lagrangian density (\ref{Lagran}), and writing $V$ 
in terms of $W$, we obtain the energy expression for a static field
\be
E = \int_{-\infty}^{\infty} \left( \half\left(\frac{d\phi}{dx}\right)^2
+  \half\left(\frac{dW}{d\phi}\right)^2 \right) dx \,.
\ee
Using the Bogomolny rearrangement \cite{Bo} (completing the square and
integrating the cross term), the energy can be reexpressed as
\be
E = \int_{-\infty}^{\infty} \half\left( \frac{d\phi}{dx}
-  \frac{dW}{d\phi} \right)^2 dx + W(1) - W(0) \,,
\ee
where the field is assumed to interpolate between $\phi = 0$ as
$x \to -\infty$ and $\phi = 1$ as $x \to \infty$. The kink minimises
the energy (for the given boundary conditions) by satisfying the
Bogomolny equation
\be
\frac{d\phi}{dx} = \frac{dW}{d\phi} = (1 - \phi^2)\phi^2 \,,
\label{Bogoeq}
\ee
and its energy is 
\be
E = W(1) - W(0) = \frac{2}{15} \,.
\ee
The Bogomolny equation can be rewritten as
\be
\left( \frac{1}{2(1 - \phi)} + \frac{1}{2(1 + \phi)} +
  \frac{1}{\phi^2} \right)d\phi = dx \,,
\ee
whose implicit solution is \cite{Loh}
\be
\half\log\frac{1+\phi}{1-\phi} - \frac{1}{\phi} = x - A \,,
\label{exactkink}
\ee
where $A$ is a constant parameter. Inverting, we obtain the kink
solution $\phi(x-A)$, which is shown in Fig.1 for $A=0$. We 
call $A$ the location of the 
kink. Another interesting position in the kink is where 
the potential $V$ has its maximum value. This is where 
$\phi = \frac{1}{\sqrt{2}}$, because $\frac{d^2W}{d\phi^2}$ is 
zero here. We call this the centre of the kink. In the present 
example it occurs at $x_{\rm centre} = A + \log(1 + \sqrt{2}) - \sqrt{2} 
\simeq A - 0.533$. This is rather awkward to deal with algebraically, 
so we usually work with the location $A$. The notion of kink centre is 
further clarified in Appendix A. Nothing physical depends on the 
distinction between location and centre.  

\begin{figure}[ht]
\begin{center}
\leavevmode
\vskip -0cm
\includegraphics[width=\textwidth]{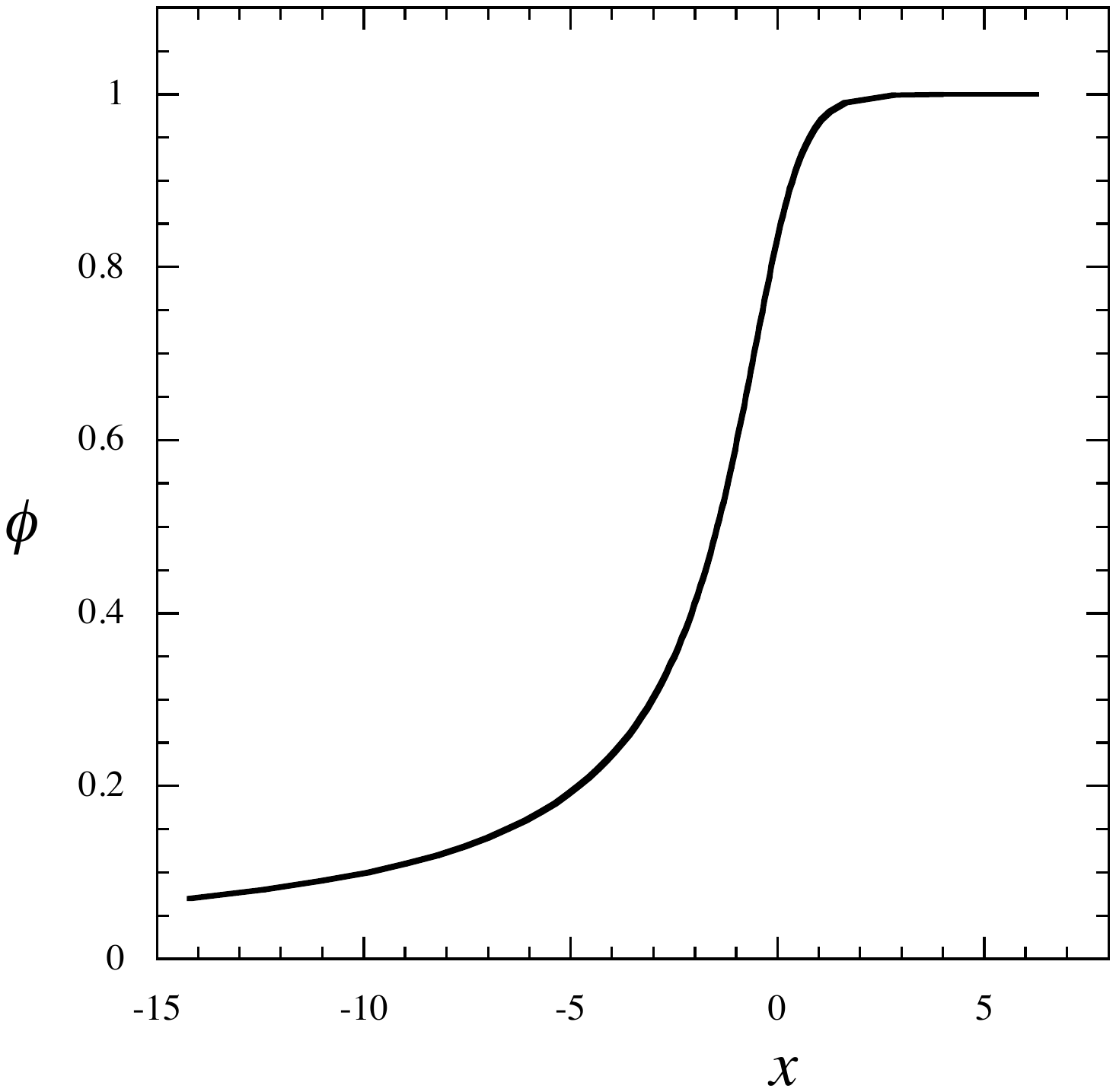}
\caption{Kink located at $A=0$.}
\vskip 0cm
\end{center} 
\end{figure}

We need to know the tail behaviours of the kink. On the left, where 
$\phi$ is near zero, the Bogomolny equation simplifies to 
$\frac{d\phi}{dx} = \phi^2$, with tail solution
\be
\phi = \frac{1}{A - x} \,.
\ee
$A$ is the same constant as before. The next term in
the expansion of $\phi$ is cubic in $\frac{1}{A - x}$, which makes the
constant $A$ in the leading term unambiguous. (A small quadratic term
$\frac{\varepsilon}{(A - x)^2}$ could be interpreted as a
shift of $A$ by $-\varepsilon$ in the leading term.) $A$
is the location where the extrapolated tail field diverges. 
On the right, where $\phi = 1 - \eta$ with $\eta$ small and 
positive, the Bogomolny equation linearises to $\frac{d\eta}{dx} =
-2\eta$, with tail solution $\eta = \exp (-2(x-b))$. The constant $b$ equals
$A-1+\half\log 2 \simeq A - 0.653$. Our main interest will be in the 
long-range tail on the left.

The transition between the tail on the left and the asymptotic field value on 
the right is rather fast, so a crude approximation to the kink with location 
$A=0$ is $\phi = -\frac{1}{x}$ for $x \le -1$ and $\phi = 1$ for $x \ge -1$. 

In the Lorentz invariant theory we are considering, the energy $E$ of a
static kink is the same as its rest mass $M$. $M$ is of course the conversion
factor between force and acceleration for kinks moving
non-relativistically. We will also need to consider the energy (mass) of a kink
with its long-range tail truncated. Consider therefore the kink
with location $A$ on the half-line $X \le x < \infty$, with $X \ll
A$. Using the Bogomolny rearrangement again, we estimate that the tail
truncation reduces the energy by
\be
E_{\rm tail} = W(\phi(X)) - W(0) = \frac{1}{3(A-X)^3} \,,
\label{tailenergy}
\ee
where we have just retained the leading terms in the kink solution, and in
$W$.

When $\phi(x-A)$ is the kink solution located at $A$, $-\phi(-x-A)$ is 
the mirror kink located at $-A$. Note that the mirror 
kink obeys the same Bogomolny equation (\ref{Bogoeq}) as the kink.
This is rather unusual, and is a consequence of $W$ having a cubic
stationary point at $\phi = 0$. The equation still has no static solution 
with both a mirror kink and kink, interpolating between $\phi = -1$ 
and $\phi = 1$. This is because the Bogomolny equation is a gradient flow
equation for the superpotential $W$, and solutions cannot pass through
stationary points of $W$. The antikink, and the antimirror-kink, obey
the Bogomolny equation with reversed sign, $\frac{d\phi}{dx} = 
-\frac{dW}{d\phi} = -(1 - \phi^2)\phi^2$. Since kink and antikink obey
Bogomolny equations with opposite signs, there are no static
kink-antikink solutions.

We conclude this section with some remarks about the well-known
mechanical reinterpretation of a kink solution. The equation for a
static field, obtained from the Lagrangian density (\ref{Lagran}), is
\be
\frac{d^2\phi}{dx^2} = \frac{dV}{d\phi} \,.
\ee
This can be interpreted as the Newtonian equation of motion for a unit mass
particle with ``position'' $\phi$ moving in ``time'' $x$ in the
inverted ``potential'' $-V$. For our kink, the particle falls
off the ``potential'' maximum at $\phi = 0$ and 
eventually stops at the  ``potential'' maximum at $\phi = 1$. Because the
total ``energy'' is zero, the motion also obeys the first-order 
Bogomolny equation. The motion away from the maximum at $\phi = 0$ is
particularly slow, because that maximum is quartic, leading to
the long-range kink tail, but the approach to the maximum at $\phi =
1$ occurs more rapidly. Below, we will consider an accelerating kink
(in the true time $t$) and will find that it approximately satisfies a
modified static equation. In the mechanical reinterpretation, this is
a Newtonian equation of motion with friction, so to get a solution for which 
the particle eventually stops at $\phi = 1$, the particle needs to leave
$\phi = 0$ with a positive ``velocity'' at a finite ``time''. Note 
that the friction is operative mainly during
the long, slow ``descent'' of the particle away from $\phi = 0$, and is
negligible during the rapid ``ascent'' to $\phi = 1$. The frictional
``force'' is comparable during these stages, since the ``velocities''
are similar, but the ``times'' over which it acts are very different. 

\vspace{5mm}

\section{Estimating Forces using Static Interpolating Fields}
\vspace{2mm}
\subsection{The Force between Kink and Mirror Kink}
\vspace{4mm}

We expect a kink and mirror kink to repel each other, and shall 
estimate their accelerations when the kink is located at $A$ and the 
mirror kink is at $-A$, for $A \gg 0$. The field may be assumed to be 
antisymmetric in $x$ at all times.

We need to produce a sensible interpolating field between 
$\phi = -1$ and $\phi = 1$. Simply adding the kink and mirror kink 
fields is not a good idea, as already mentioned.
Instead we split up the spatial line at points $-X$ and $X$, with $0 \ll
X \ll A$, so that for $x \le -X$ we have an exact mirror kink field, and
for $x \ge X$ an exact kink field. In between, for $-X \le x \le X$, we 
assume the interpolating field has the linear behaviour $\phi(x) = \mu
x$. This is justified by the linearised, static field equation for small
$\phi$, which is simply $\frac{d^2\phi}{dx^2} = 0$.   
We require the field and its first spatial derivative to be continuous
at $X$ (and also $-X$). This leads to the conditions
\be
X = \half A \quad {\rm and} \quad \mu = \frac{4}{A^2} \,,
\label{Xmu}
\ee
where we have assumed the tail formula for the kink field, which 
is approximately valid at $X$. The interpolating field for $A = 10$
is shown in Fig.2.

\begin{figure}[ht]
\begin{center}
\leavevmode
\vskip -0cm
\includegraphics[width=\textwidth]{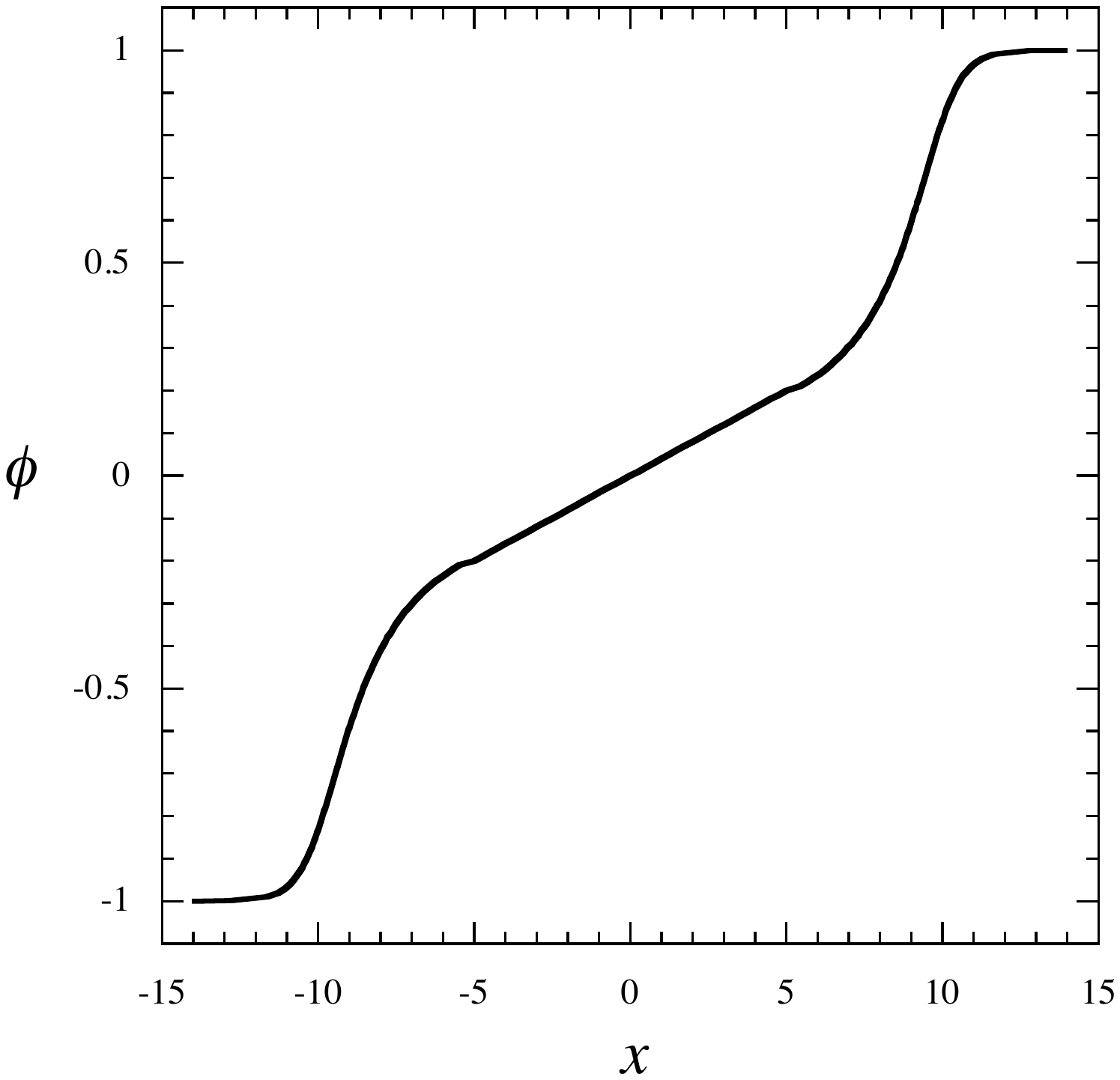}
\caption{Interpolating field for kink at $A=10$ and mirror 
kink at $A=-10$, with linear interpolation between $x=-5$ and $x=5$. }
\vskip 0cm
\end{center} 
\end{figure}

Note that if we had solved the static field equation exactly between
the mirror kink and kink, the appropriate continuity conditions would 
have been impossible to solve, as they would have given us a global 
smooth, static solution, which doesn't exist. This impasse is resolved by
using an approximate solution, as we have done.

Let us now calculate the total energy of this interpolating field, to
leading order in $\frac{1}{A}$. This is the sum of the energies of the
kink and mirror kink, with their tails truncated, and the energy of the linear
interpolating field. We calculated the tail energy in
eq.(\ref{tailenergy}); here it is $\frac{8}{3A^3}$. For the 
linear part of the field, the energy is
\be
E_{\rm lin.} = \int_{-X}^{X} \left( \half\left(\frac{d\phi}{dx}\right)^2
+  \half\left(\frac{dW}{d\phi}\right)^2 \right) dx \,.
\ee 
We set $\phi = \mu x$ and make the approximation
$\frac{dW}{d\phi} = \phi^2 = \mu^2 x^2$. The integral gives $\mu^2 X +
\frac{1}{5} \mu^4 X^5$. Now we use the values (\ref{Xmu}) and find that
$\mu^2 X$ and $\mu^4 X^5$ are both of the same order in $\frac{1}{A}$
as the tail energy, and the total energy is
\be
E = \frac{4}{15} - \frac{16}{3 A^3} + \frac{8}{A^3} + \frac{8}{5 A^3}
= \frac{4}{15} + \frac{64}{15 A^3} \,.
\ee 
Using this as the potential energy, and dropping the constant, we
obtain an effective Lagrangian for the motion of the two kinks at
large separation
\be
L_{\rm kinks} = \frac{2}{15} {\dot A}^2 - \frac{64}{15 A^3} \,,
\ee
where the kinetic energy is twice that of a single, freely moving
kink. The equation of motion is
\be
{\ddot A} = \frac{48}{A^4} \,,
\label{accel1}
\ee
showing that the kink has acceleration $\frac{48}{A^4}$ to the
right. The mirror kink has opposite acceleration, so the kink and 
mirror kink repel each other. The force acting is the mass 
$M = \frac{2}{15}$ times the acceleration, and is   
\be
F = \frac{32}{5 A^4} \,.
\ee
Alternatively, $F$ is the negative of the derivative of $E$ with respect to the
separation of the kink and mirror kink. The separation is $2A$ with 
an ambiguity of order $1$ because of the asymmetry of the kinks and
the finite widths of their cores, and because the notion of kink
location or centre is to some extent a matter of definition, but 
this ambiguity does not affect a force falling off as a power of
$\frac{1}{A}$, nor its coefficient, to leading order.

This calculation gives the expected dependence of $F$ on $A$, but
the coefficient is not correct. This is because the coefficient is
sensitive to the interpolation used. Also, the force is not acting
uniformly throughout each kink, but only in the region of the linear
interpolating field. One knows this because in the regions of the
exact, static kink and mirror kink solutions, no force 
is acting. 

\subsection{The Force between Kink and Antikink}
\vspace{4mm}

A well-separated kink at $A$ and an antikink at $-A$ are expected to 
attract, and the field can be assumed to be symmetric in $x$ at all
times. Again, we split up the 
spatial line at $-\half A$ and $\half A$, so that for $x \le -\half A$ 
we have an exact antikink field, and for $x \ge \half A$ an exact kink 
field. In between, for $-\half A \le x \le \half A$, we assume the 
interpolating field has quadratic behaviour $\phi(x) = 
\alpha + \beta x^2$, and we fix the values of $\alpha$ and $\beta$ 
so that $\phi$ is continuous and has continuous first derivative 
at $\pm\half A$. These continuity conditions are satisfied if 
\be
\alpha = \frac{1}{A} \quad {\rm and} \quad \beta = \frac{4}{A^3} \,,
\label{alphabeta}
\ee
where we have assumed the tail formula for the kink field, which 
is approximately valid at $\half A$. The splitting points are the 
same as those used in the kink-kink case. There could be a better 
choice, $-X$ and $X$ with $0 \ll X \ll A$ and $X \ne \half A$, but 
the algebra gets more complicated. The interpolating field is shown in
Fig.3.

\begin{figure}[ht]
\begin{center}
\leavevmode
\vskip -0cm
\includegraphics[width=\textwidth]{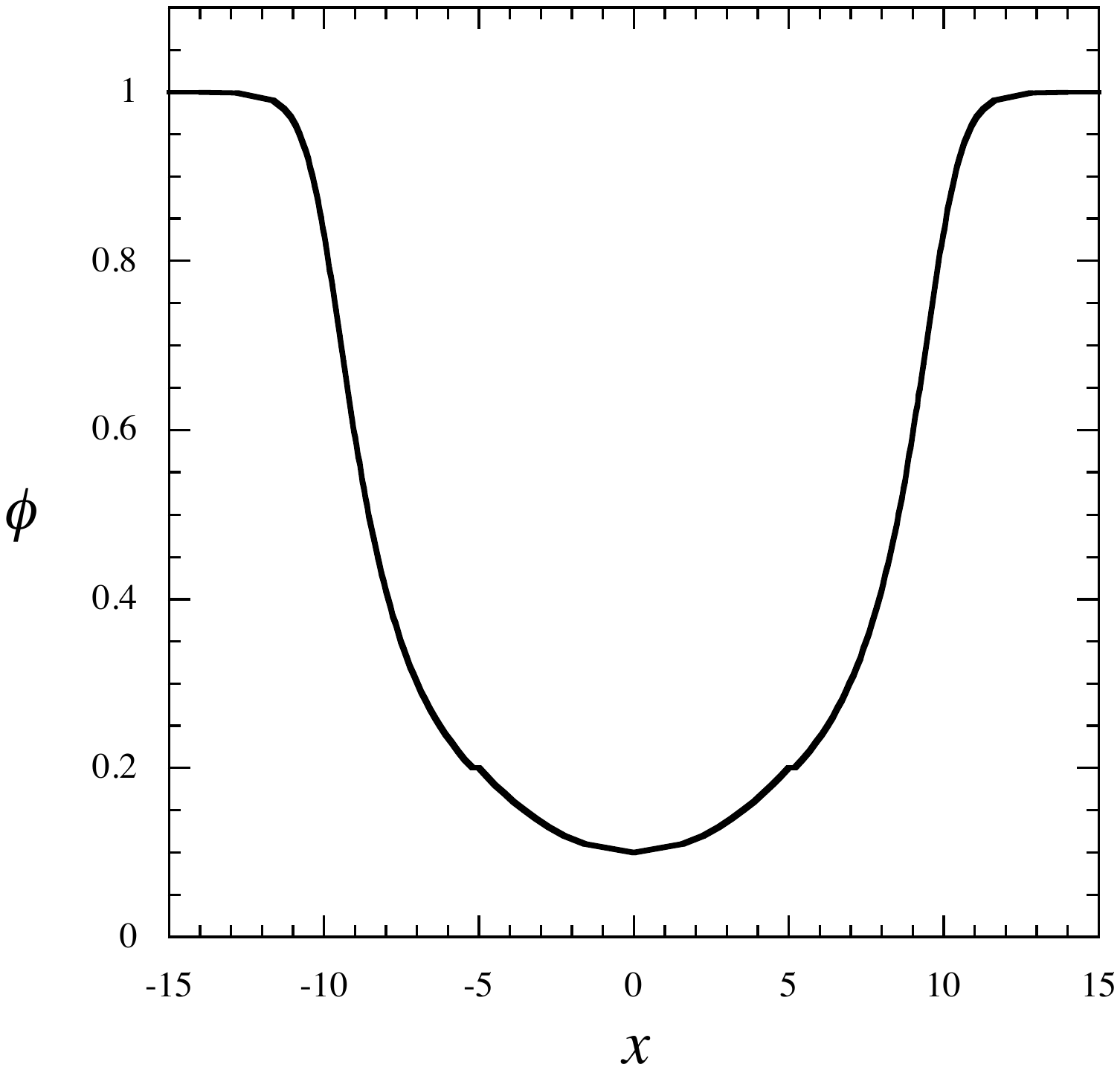}
\caption{Interpolating field for kink at $A=10$ and antikink at
  $A=-10$, with quadratic interpolation between $x=-5$ and $x=5$.}
\vskip 0cm
\end{center} 
\end{figure}

The total energy of this interpolating field, to leading order in 
$\frac{1}{A}$, is the sum of the energies of the truncated kink and antikink, 
with their tail energies $\frac{8}{3A^3}$ removed, and the energy of 
the quadratic interpolating field
\be
E_{\rm quad.} = \int_{-\half A}^{\half A} 
\left( \half\left(\frac{d\phi}{dx}\right)^2
+  \half\phi^4 \right) dx \,,
\ee 
where $\phi = \frac{1}{A} + \frac{4}{A^3}x^2$ and we have again made 
the approximation $\frac{dW}{d\phi} = \phi^2$ as $\phi$ is small. The 
integral is elementary, but a bit more elaborate than for the 
linear interpolation in the kink-kink case. We find 
$E_{\rm quad.} = \frac{1504}{315 A^3}$, so the total energy is
\be
E = \frac{4}{15} - \frac{16}{3 A^3} + \frac{1504}{315 A^3} =
\frac{4}{15} - \frac{176}{315 A^3} \,.
\ee 
The estimated force between the kink and antikink is minus the 
derivative of $E$ with respect to the separation $2A$,
\be
F = -\frac{88}{105 A^4} \,.
\ee
It decays with the fourth power of the separation and is attractive. 
As the kink mass is $M = \frac{2}{15}$, the acceleration of the kink is
\be
{\ddot A} = -\frac{44}{7A^4} \,.
\label{accel2}
\ee

The strength of the kink-kink repulsive force (using the same 
approximate approach) was $\frac{32}{5A^4}$, so the strength of the
attraction in the kink-antikink case appears to be about one eighth of
this. However, these calculations are sensitive to the choice of 
splitting points and interpolating fields, so the numerical
coefficients are not very reliable. In the next section we will 
use a different approach, where the kinks are accelerating 
and slightly deformed, and will derive more reliable forces.

\vspace{5mm}

\section{Accelerating Kinks and Antikinks}
\vspace{2mm}
\subsection{Kink and Mirror Kink}
\vspace{4mm}

In this approach to the force, we model the field $\phi(x,t)$, 
assuming that the kink and mirror kink are located at 
well-separated, time-dependent positions $A(t)$ and $-A(t)$.  
The idea is to find an implicit profile for the kink with acceleration
$a = {\ddot A}$ that describes the field for $x>0$, and at least
approximately solves the field equation. The mirror kink has
the reflected field (sign-reversed $x$, $\phi$ and $a$) for 
$x<0$. The accelerating kink has a distorted profile, whose tail
is no longer $\frac{1}{A-x}$. Instead, $\phi$ is zero at some
finite distance to the left of the kink, the distance depending on the 
acceleration. The kink profile is linear near here, so it can be 
continuously glued to the profile of the accelerating mirror kink. 
Gluing at $x=0$ creates a field antisymmetric in
$x$, with a continuous first derivative.

This interpolation does not give an exact solution, because the
acceleration jumps discontinuously from $a$ to $-a$ at $x=0$, and
because we need to make various approximations that are explained in
more detail below. Tail gluing nevertheless creates a convenient interpolation.

We introduce here a result related to Noether's theorem for
conserved momentum. By a standard argument \cite{book}, the momentum
density in the field theory we are considering is $-\frac{\pr\phi}
{\pr t}\frac{\pr\phi}{\pr x}$, and its integral over the whole spatial line  
is conserved provided $\phi$ satisfies its field equation. The time
derivative of the momentum integral $P$ over a finite interval is a 
sum of endpoint (surface) terms. In particular, using the field 
equation, one finds that for a field on the half-line $x \ge X$, 
obeying the kink boundary condition $\phi \to 1$ as $x \to \infty$,  
\be
\frac{dP}{dt} = -\frac{d}{dt} \int_X^\infty \frac{\pr\phi}{\pr t}
\frac{\pr\phi}{\pr x} \, dx 
= \left(\half\left(\frac{\pr\phi}{\pr t}\right)^2 
+ \half\left(\frac{\pr\phi}{\pr x}\right)^2 
- V(\phi(x))\right) \Bigg|_{x=X} \,.
\label{force}
\ee
For the fields we are interested in, the term 
$\half\left(\frac{\pr\phi}{\pr t}\right)^2$ is negligible.
One may interpret $\frac{dP}{dt}$ as the 
force $F$ acting on the entire field to the right of $X$, and it
appears to be exerted at the point $X$, although we shall refine 
this interpretation later. For the field of the accelerating kink and
mirror kink we can choose $X=0$, and there $\phi = 0$ and $V = 0$. The
force on the kink becomes
\be
F = \half\left(\frac{\pr\phi}{\pr x}\right)^2 \Bigg|_{x=0} \,.
\label{Forceslope}
\ee

The last formula gives a rough estimate of the force,
even when one does not attempt to solve the field equation. For
example, if we simply add the static tail fields of the kink and 
mirror kink, then near $x=0$,
\be
\phi = \frac{1}{A-x} - \frac{1}{A+x} \,,
\label{tailsum}
\ee
so $\frac{d\phi}{dx} = \frac{2}{A^2}$ at $x=0$, giving the force estimate
$F = \frac{2}{A^4}$, and the kink acceleration estimate
${\ddot A} = \frac{15}{A^4}$. This has the right
quartic dependence on $\frac{1}{A}$, but the coefficient does not
agree with what we found earlier, nor with what we claim is the more
accurate value determined below. 

Now let's return to finding an approximate solution of the field equation
representing the accelerating kink. We do not need to
know that the acceleration is produced by the mirror kink to the 
left. We model the kink by a field of the form
\be
\phi(x,t) = \chi(x - A(t)) \,.
\label{accelfield}
\ee
The acceleration is $a = {\ddot A}$, which we assume is small. We
also assume the squared velocity ${\dot A}^2$ remains much less 
than $1$, so the motion is non-relativistic, and Lorentz contraction of
the kink and radiation can be ignored. Throughout, we work to linear 
order in $a$. It is convenient to denote the argument of $\chi$ by 
$y$ and denote a derivative of $\chi$ with respect to $y$ by $'$.

The first time derivative of $\phi$ is $-{\dot A}\chi'$, and we approximate   
the second time derivative by $-{\ddot A}\chi' = -a\chi'$, dropping the term
proportional to ${\dot A}^2$. (The term 
$\half\left(\frac{\pr\phi}{\pr t}\right)^2$ on the right 
hand side of eq.(\ref{force}) is similarly
proportional to ${\dot A}^2$, which justifies its neglect.) 
Substituting the accelerating field (\ref{accelfield}) into the 
field equation (\ref{fieldeq}) then gives
\be
\chi'' + a\chi' - \frac{dV(\chi)}{d\chi} = 0 \,.
\label{chieq}
\ee
The profile $\chi$ satisfies this static equation with $a$ as parameter, 
so it evolves adiabatically with time as $a$ varies. $\phi$ evolves both
because of this evolution of $\chi$, and because $A(t)$ occurs in the 
argument of $\chi$. 

Using the mechanical analogy discussed in Section 2, eq.(\ref{chieq}) is
interpreted as the motion of a ``particle'' with ``position'' $\chi$ in the 
inverted potential $-V$, now subject to friction with a friction 
coefficient $a$.
The kink boundary condition is $\chi \to 1$ as $y \to \infty$.
The potential $-V$ is zero at both $\chi = 0$ and $\chi = 1$, so for the
``particle'' to reach $\chi = 1$ eventually and stop there, it must leave
$\chi = 0$ with a finite ``velocity'' at a finite, but arbitrary
time. This ``velocity'' is determined by $a$. 

Reverting back to field language, we anticipate a solution of 
eq.(\ref{chieq}) where $\chi = 0$ and $\chi'$ is positive at some finite $y$, 
and $\chi \to 1$ as $y \to \infty$. The equation for $\chi$ is
translation invariant, so any solution can be shifted to the left or
right. The solution we require is the one where $\chi(x-A)$ agrees as
closely as possible with the static kink solution $\phi(x-A)$ in the core
region of the kink. 

This requirement is not so easy to implement. The short-range 
tails to the right have slightly different exponential forms, so their
coefficients cannot really be matched. Instead, one could require, for example,
that $\phi$ and $\chi$ take identical values $\frac{1}{\sqrt{2}}$ at
the same position. This could be implemented numerically, but it
cannot be done analytically because we do not have even an implicit
solution for $\chi$. 

The approach we have adopted to this problem is to match the long-range 
tails of $\phi$ and $\chi$ close to the core regions, and 
in particular to arrange that the positions where the extrapolated 
long-range tails diverge are the same. This assumes that the main 
effect of the friction term is on these tails, and its effect in the 
kink core and further to the right is negligible, as we argued earlier.

For the static kink, the extrapolated tail $\frac{1}{A-x}$ diverges at $x =
A$. For $\chi$ we calculate as follows. In the small term $a\chi'$ in 
eq.(\ref{chieq}) we can assume that $\chi$ is the undeformed kink, for 
which $\chi' = \frac{dW}{d\chi}$. We can therefore trade the friction 
term for a modified potential $\widetilde{V} = V - aW$ and obtain a 
first integral of eq.(\ref{chieq}) of the form
\be
\chi'^2 = 2\widetilde{V} + {\rm const.}
\label{modeqchi}
\ee
The constant is zero, because $\chi'$, $V$ and $W$ are all zero as 
$y \to \infty$. In the long-range tail region we can now make the 
approximations $V(\chi) = \half\chi^4$ and $aW = -\frac{2a}{15}$, 
obtaining from eq.(\ref{modeqchi}) the simplified equation
\be
\frac{d\chi}{dy} = \sqrt{\frac{4a}{15} + \chi^4} \,.
\label{simplechi}
\ee
As a consistency check, note that the linear behaviour of $\chi$ near
its zero has slope $\mu = \sqrt{\frac{4a}{15}}$, so the force on the
kink, according to eq.(\ref{force}), is $\frac{2a}{15}$. This is the
product of the kink's mass and acceleration, as it should be.  

The solution of eq.(\ref{simplechi}) still involves an
elliptic integral of the first kind, but fortunately we just need the definite
integral. The solution $\chi(y)$ should have the properties $\chi(-A)
= 0$ and $\chi(0)$ diverges. Then $\chi(x-A)$ will be zero at
$x=0$ and $\chi(x-A)$ will diverge at $x=A$, as we require for the 
extrapolated tail field. Therefore
\be
\int_0^\infty \frac{d\chi}{\sqrt{\frac{4a}{15} + \chi^4}} = A \,.
\ee
After changing variable to $\chi = \left(\frac{4a}{15}\right)^{\quart}
\lambda$, this relation simplifies to
\be
I \equiv \int_0^\infty \frac{d\lambda}{\sqrt{1 + \lambda^4}} =
\left(\frac{4a}{15}\right)^{\quart} A \,.
\ee
$I$ is a complete elliptic integral, and the further
change of variable $\lambda^2 = \sinh\nu$ gives the equivalent relation
\be
I = \half\int_0^{\infty} (\sinh\nu)^{-\half} \, d\nu =
\left(\frac{4a}{15}\right)^{\quart} A \,.
\ee
Using either form of the integral $I$ \cite{GR,WWW} we find that
\be
a = {\ddot A} = \frac{15}{4} 
\left(\frac{\Gamma\left(\frac{1}{4}\right)^2}{4\sqrt{\pi}}\right)^4
\frac{1}{A^4} \simeq \frac{15}{4}(1.854)^4 \frac{1}{A^4} =
\frac{44.3}{A^4} \,.
\ee
This is our result for the acceleration. The coefficient $44.3$ is
different from the previous estimates of $48$ and $15$. The force 
on the kink is
\be
F = \frac{2}{15}{\ddot A} \simeq \half(1.854)^4 \frac{1}{A^4}
= \frac{5.91}{A^4} \,.
\ee
We can go back through the calculation, and verify that terms that were
dropped, e.g. $\frac{1}{3}a\chi^3$ in $aW$, are small compared to 
those retained. 

The equation of motion for the kink, ${\ddot A} = \frac{44.3}{A^4}$,
implies the energy conservation equation 
\be
\half {\dot A}^2 + \frac{1}{3}\frac{44.3}{A^3} = {\rm const.}
\ee
and from this we deduce that if the kink and mirror kink approach from 
infinity with a small speed $v$, then at closest approach
\be
A = \left(\frac{29.5}{v^2}\right)^{\frac{1}{3}} \,.
\ee

This calculation, involving an accelerating kink, seems to give a
more reliable result than the earlier, static approach because 
the force is properly distributed along the kink. 
To see this, we multiply eq.(\ref{chieq}) by $\chi'$ and integrate
from $X$ to $\infty$, for $X \ge 0$, obtaining
\be
a\int_X^{\infty} \chi'^2 \, dy = \left( \half\chi'^2 -
  V(\chi(y))\right) \Bigg|_{y=X} \,.
\ee
By analogy with eq.(\ref{force}), the right hand side is the force
acting on the half-line to the right of $X$. Since $a$ is small, on
the left hand side we can replace $\chi'$ by the derivative of the
static kink. The Bogomolny equation then implies that $\chi'^2$ is the 
energy density of the static kink, to the accuracy we need. (Of course, we
can't use the Bogomolny equation on the right hand side, because that 
would give zero.) Therefore the left hand side is the mass on the 
half-line to the right of $X$, times the acceleration 
$a$. As $X$ is not fixed, we deduce that the force is everywhere of 
the correct strength.

We have apparently lost track of the mirror kink, but this occupies
the half-line $x \le 0$, and its field is
$-\phi(-x,t) = -\chi(-x-A(t))$. The 
acceleration of the mirror kink is $-a$. The kink and mirror kink
fields meet at $x=0$ and have a continuous derivative. The mirror kink
satisfies an equation like (\ref{chieq}) but with the sign of $a$
reversed. The second spatial derivative of the field $\phi$ at $x=0$ 
therefore has a discontinuity $2a\chi'$, which we see from 
eq.(\ref{simplechi}) is of order $a^{\frac{3}{2}}$. This discontinuity
is presumably negligible, but is a consequence of the several 
approximations we have made. An even better approximation
would involve a smoother interpolation of the acceleration through $x=0$.

\subsection{Kink and Antikink}

Consider next a kink and antikink located at $A(t)$ and $-A(t)$, 
with $A \gg 0$. The profile of the accelerating kink now has a tail 
whose spatial derivative is expected to be zero at some finite 
position to the left of the kink. We arrange that this position is at
$x=0$, which gives a relation between the acceleration and kink 
location $A$. The antikink has the reflected field (sign-reversed $x$) for 
$x \le 0$. At $x=0$ we can glue the kink and antikink profiles together
to create a field symmetric in $x$, which is continuous and has 
continuous first and second spatial derivatives. For similar reasons as in the
kink-kink case, this approach does not give an exact solution.

Recall the Noether formula (\ref{force}) for the rate of 
change of field momentum $P$ on the half-line $x \ge X$ (with $X \ge 0$), 
interpreted as the force $F$ acting on the field to the right of 
$X$. While the kink and antikink are slowly moving, the squared 
time-derivative term can again be neglected. We choose $X=0$, and as 
$\frac{\pr\phi}{\pr x} = 0$ at $x=0$ the force on the kink becomes
\be
F = -V(\phi(0)) \,.
\label{ForceV}
\ee

This estimates the force even in the
absence of a proper solution of the field equation. For example, 
$\phi(0) = \frac{1}{A}$ for the static interpolating field we 
constructed in Section 3, and if we use the approximation 
$V(\phi) = \half\phi^4$ then $F = -\frac{1}{2A^4}$, so 
${\ddot A} = -\frac{15}{4A^4}$. This has the expected sign 
and quartic dependence on $\frac{1}{A}$. The coefficient 
is similar to what we found earlier, using the energy of the static field. 

As in the kink-kink case, we model the accelerating kink by a field
$\phi(x,t) = \chi(x - A(t))$, and let $y$ denote the argument of $\chi$.
Because of the expected attraction to the antikink, we assume ${\ddot A}$ 
to be negative, and introduce $a = -{\ddot A}$, with $a$ positive and 
small. As before, ${\dot A}^2$ is supposed small enough to be neglected.
The first time derivative of $\phi$ is $-{\dot A}\chi'$, and we approximate   
the second time derivative by $-{\ddot A}\chi' = a\chi'$, dropping the 
${\dot A}^2$ term. Substituting the accelerating field 
into the field equation (\ref{fieldeq}) then gives
\be
\chi'' - a\chi' - \frac{dV(\chi)}{d\chi} = 0 \,,
\label{chieq2}
\ee
the same equation as (\ref{chieq}) but with reversed sign of $a$. 

In the mechanical analogy, eq.(\ref{chieq2}) describes the motion of 
a ``particle'' subject to negative friction with coefficient $a$. 
For the ``particle'' to reach $\chi = 1$ eventually and 
stop there, it now needs to depart at zero ``velocity'' at some 
finite ``time'' from a positive ``position'' $\chi$ where $-V$ is 
negative. The initial ``energy'' is then negative, but increases 
with the ``time'' $y$.  

In field language, we anticipate a solution of eq.(\ref{chieq2}) 
where $\chi$ is positive and $\chi' = 0$ at some finite $y$, and which 
satisfies the kink boundary condition $\chi \to 1$ as $y \to \infty$. The 
required solution, with $\chi(x-A)$ matching the static kink 
$\phi(x-A)$ in the kink core, is again found by arranging that 
the extrapolated long-range tails diverge at the same location $A$.

The calculation of the tail of $\chi$ differs 
now that the term proportional to $a$ in eq.(\ref{chieq2}) has the 
opposite sign. Replacing $\chi'$ by $\frac{dW}{d\chi}$ in this 
term leads to the modified potential $\widetilde{V} = V + aW$ and 
the first integral
\be
\chi'^2 = 2\widetilde{V} \,.
\label{modeqchi2}
\ee
In the long-range tail region we make the 
approximations $V(\chi) = \half\chi^4$ and $aW = -\frac{2a}{15}$, 
obtaining from eq.(\ref{modeqchi2})
\be
\frac{d\chi}{dy} = \sqrt{\chi^4 -\frac{4a}{15}} \,.
\label{simplechi2}
\ee
The range of $\chi$, allowing for the divergence of the extrapolated 
tail, is from $\left(\frac{4a}{15}\right)^{\quart}$ to $\infty$. At 
the lower end of this range, $\frac{d\chi}{dy} = 0$ and $
V(\chi) = \frac{2a}{15}$, which implies that the force (\ref{ForceV}) 
is $-\frac{2a}{15}$, the product of the kink's mass and 
acceleration. We can therefore consistently require that 
$\chi = \left(\frac{4a}{15}\right)^{\quart}$ at $x=0$.  

As before, we just need the definite integral of eq.(\ref{simplechi2}). The 
solution $\chi(y)$ should have the properties $\chi(-A)
= \left(\frac{4a}{15}\right)^{\quart}$ and $\chi(0)$ diverges. Then 
$\chi(x-A) = \left(\frac{4a}{15}\right)^{\quart}$ at $x=0$ and 
$\chi(x-A)$ diverges at $x=A$, as required. Therefore
\be
\int_{\left(\frac{4a}{15}\right)^{\quart}}^\infty 
\frac{d\chi}{\sqrt{\chi^4 - \frac{4a}{15}}} = A \,.
\ee
After rescaling, this simplifies to
\be
J \equiv \int_1^\infty \frac{d\lambda}{\sqrt{\lambda^4 -1}} =
\left(\frac{4a}{15}\right)^{\quart} A \,.
\ee
$J$, like the integral $I$, is a complete elliptic integral, and the further
change of variable $\lambda^2 = \cosh\nu$ gives
\be
J = \half\int_0^{\infty} (\cosh\nu)^{-\half} \, d\nu =
\left(\frac{4a}{15}\right)^{\quart} A \,.
\ee
Knowing the integral $J$ \cite{GR,WWW}, we find
\be
\left(\frac{4a}{15}\right)^{\frac{1}{4}} A = \frac{1}{\sqrt{2}}
\frac{\Gamma\left(\frac{1}{4}\right)^2}{4\sqrt{\pi}} \,.
\ee
The acceleration of the kink is therefore
\be
{\ddot A} = -a = -\frac{15}{16} 
\left(\frac{\Gamma\left(\frac{1}{4}\right)^2}{4\sqrt{\pi}}\right)^4
\frac{1}{A^4} \simeq
-\frac{11.1}{A^4} \,.
\ee
The coefficient $11.1$ differs from the previous estimates 
of $\frac{44}{7}$ and $\frac{15}{4}$ for the kink-antikink case. 
The force on the kink is
\be
F = \frac{2}{15}{\ddot A} \simeq -\frac{1.48}{A^4} \,,
\ee 
confirming that the kink and antikink attract. 

The antikink occupying the half-line $x \le 0$ has the 
field $\phi(-x,t) = \chi(-x-A(t))$, and the antikink's acceleration is $a$. The 
kink and antikink fields meet at $x=0$ with zero spatial derivative. 
The antikink satisfies an equation like (\ref{chieq}) but with the sign of $a$
reversed. Therefore, the second spatial derivative of the field 
$\phi$ at $x=0$ is continuous, because the first derivative is zero
there, but curiously, the third derivative has a small discontinuity.

The most interesting aspect of the kink-antikink force is the extra factor of 
$\quart$ (and the reversed sign) compared with the force between 
a kink and mirror kink. This arises from the extra factor of
$\frac{1}{\sqrt{2}}$ in the integral $J$, compared to $I$. One can 
understand this factor using a contour integral argument, without 
evaluating the integrals.\footnote{I am grateful to Joe Davighi and 
Alex Abbott for this insight.} Recall that
\be
I = \int_0^\infty \frac{d\lambda}{\sqrt{1 + \lambda^4}} \,, \quad 
J = \int_1^\infty \frac{d\lambda}{\sqrt{\lambda^4 -1}} \,,
\ee
and note that by a change of variable 
\be
J = \int_0^1 \frac{d\lambda}{\sqrt{1 - \lambda^4}} \,.
\ee
By considering the integral of $\frac{1}{\sqrt{1 + z^4}}$ along
the contour from $0$ to $\infty$ to $\infty e^{\frac{\pi i}{4}}$ and
back to $0$ via the branch point at $z = e^{\frac{\pi i}{4}}$, we
find that
\be
I + e^{\frac{3\pi i}{4}} J - e^{\frac{\pi i}{4}} J = 0 \,,
\ee
so $J = \frac{1}{\sqrt{2}} I$.

\vspace{5mm}

\section{Conclusions}
\vspace{4mm}

We have investigated a simple example of a kink with a long-range
tail in scalar field theory. The tail has $\frac{1}{x}$ behaviour, 
because the potential $V$ in the field theory has a quartic minimum. We 
have derived the repulsive force between a kink and a mirror kink, and
the attractive force between a kink and an antikink, when their long-range 
tails overlap. The forces are proportional to the inverse fourth power 
of the separation when the separation is large. The numerical 
coefficients have been calculated by allowing for the kink 
accelerations and then solving modified Bogomolny equations. We have
argued that this approach is likely to give the most reliable
coefficients multiplying the power of the separation. The 
coefficients are transcendental. 

Interestingly, the strength of the 
repulsion between kink and mirror kink is four times the attraction 
between kink and antikink, contrasting with the equal strengths of the
kink-kink repulsion and kink-antikink attraction for kinks with
short-range tails. This seems to be a consequence of the 
long-range tails being solutions of a nonlinear equation of type
$\frac{d^2\phi}{dx^2} = 2\phi^3$ when $V$ has a quartic minimum, so
the tails cannot simply be superposed, but it would be valuable to
find a robust proof. It should not be difficult to 
generalise our discussion to variant kinks with long-range
tails, and investigate this phenomenon. It would also be interesting 
to solve the equation of motion for incoming kinks with small 
velocities, using the forces we have obtained, and compare with a 
numerical simulation of the dynamics using the full field 
equation.\footnote{Since completion of this work, I.C. Christov 
et al. \cite{Chr2} have calculated the forces between kinks and
antikinks with long-range tails in the family of scalar field theories with
potentials $V(\phi) = \half(1 - \phi^2)^2 \phi^{2n}$, for $n = 2,3,4,\dots$. 
The forces decay with an inverse power of the separation depending
on $n$, and the ratio of the kink-kink repulsion and kink-antikink
attraction also depends on $n$. Numerical simulations of the full
field equations confirm that the approach to the force calculation 
allowing for kink and antikink accelerations, as described in Section 
4, gives reliable coefficients multiplying the inverse power of the 
separation.}

\vspace{4mm}

\section*{Acknowledgements}

I am grateful to Avadh Saxena for drawing my attention to this
problem, and for subsequent correspondence. I also thank Avadh, Emil 
Mottola, and their colleagues, for hospitality during my visit to Los 
Alamos in August 2018. I thank Yasha Shnir for helpful
correspondence, and Anneli Aitta for producing the figures. This 
work has been partially supported by STFC consolidated grant ST/P000681/1.

\vspace{5mm}

\section*{Appendix A: Kink Centre}
\vspace{4mm}

Let $\phi_n$ and $\phi_{n+1}$ be adjacent zeros of $\frac{dW}{d\phi}$,
and hence of the potential $V = \half\left(\frac{dW}{d\phi}\right)^2$. 
We assume $\frac{dW}{d\phi}$ is positive between these zeros. Then
the Bogomolny equation $\frac{d\phi}{dx} =\frac{dW}{d\phi}$ has the 
correct sign for a kink interpolating between $\phi_n$ and
$\phi_{n+1}$, and the kink solution $\phi(x)$ increases 
monotonically with $x$.

$V$ could have several local maxima and minima between $\phi_n$ and
$\phi_{n+1}$, but let us assume that it has just one maximum, denoted 
by $\phi_{n+\half}$. For the following reasons, the position 
$x = x_{\rm centre}$ where $\phi = \phi_{n+\half}$ can be regarded 
as the centre of the kink. At $x_{\rm centre}$, 
\be
\frac{dV}{d\phi} = \frac{dW}{d\phi}\frac{d^2W}{d\phi^2} = 0 \,,
\ee
so $\frac{d^2W}{d\phi^2} = 0$, as $\frac{dW}{d\phi}$ is
non-zero. Differentiating the Bogomolny equation gives, at $x_{\rm centre}$,
\be
\frac{d^2\phi}{dx^2} = \frac{d^2W}{d\phi^2} \frac{d\phi}{dx} = 0 \,,
\ee
so $x_{\rm centre}$ is the point of inflection in the kink
profile, the position where $\phi$ is increasing with $x$ most
rapidly. The energy density of a kink satisfying the Bogomolny
equation is $2V$, so this is also maximal at $x_{\rm centre}$.
Let $\phi(x)$ be a centred kink, for which $x_{\rm centre} = 0$. The 
general kink is then $\phi(x-c)$ with centre $c$. 

For the familiar $\phi^4$ and sine-Gordon kinks, the
centres are the obvious points about which the kink is antisymmetric.
The centre of the kink with long-range tail, considered in the main
part of this paper, was clarified in Section 2.
Another non-trivial example is the $\phi^6$ kink \cite{Loh}. Here,
\be
V(\phi) = \half(1-\phi^2)^2\phi^2 \,,
\ee
with quadratic minima at $\pm 1$ and $0$. The quadratic behaviours are
different at $0$ and $1$, so the interpolating kink solution is not 
reflection (anti)symmetric. As 
$\frac{dW}{d\phi} = (1-\phi^2)\phi$, 
\be
W(\phi) = \half \phi^2 - \quart \phi^4 + {\rm const.}
\ee
and the Bogomolny equation is 
\be
\frac{d\phi}{dx} = (1-\phi^2)\phi \,.
\ee
Using partial fractions, this can be integrated to give
\be
\phi(x) = \left(1 + 2e^{-2(x-c)}\right)^{-\half} \,.
\label{phi6kink}
\ee   
The kink centre is where $\frac{d^2W}{d\phi^2} = 1 - 3\phi^2 = 0$, 
i.e. where $\phi = \frac{1}{\sqrt{3}}$. The expression (\ref{phi6kink}) has 
been carefully normalised so that $x_{\rm centre}=c$. The kink energy 
is $E = W(1) - W(0) = \quart$.  

\vspace{5mm}

\section*{Appendix B: Force between $\phi^6$ Kinks}
\vspace{4mm}

Here, for completeness, we illustrate the simple method that 
gives the force between two kinks having short-range tails. The 
following example was not explicitly considered in \cite{Ma5,book}.

The $\phi^6$ theory has the kink solution (\ref{phi6kink}) interpolating
between $\phi = 0$ and $\phi = 1$, and a mirror kink interpolating
between $\phi = -1$ and $\phi = 0$ obtained by reversing the signs 
of $x$ and $\phi$. A field with a kink centred at $c$
and mirror kink centred at $-c$, with $c \gg 0$, is well described 
by the linear superposition
\be
\phi(x) = \left(1 + 2e^{-2(x-c)}\right)^{-\half} - 
\left(1 + 2e^{2(x+c)}\right)^{-\half} \,. 
\label{phi6super}
\ee
The kinks tails are short-ranged here, i.e. exponentially decaying, so
the linear superposition is better justified than for the kinks with
long-range tails we discussed earlier. The kink obeys the Bogomolny
equation and the mirror kink the equation with reversed sign. So the 
linear superposition does not obey either Bogomolny equation, and is 
therefore not an exact static solution. However, both kink tails obey 
the linearised second-order static field equation in the region
between the kinks, so the sum of the tails does too. This explains why the
linear superposition is a good interpolating field.

To find the force exerted by the mirror kink on the kink, we use the
Noether formula for the rate of change of momentum 
(\ref{force}). At a point $X$ between the kinks, with $-c \ll X \ll
c$, the force acting on the kink to the right is
\be
F = \left(\half\left(\frac{d\phi}{dx}\right)^2 - \half\phi^2 \right)
\Bigg|_{x=X} \,,
\label{tailsforce}
\ee
where we have made the approximation $V(\phi) = \half\phi^2$, as $\phi$
is near zero. The superposed tail field, obtained from the leading 
exponentially small terms in (\ref{phi6super}), is
\be
\phi(x) = \frac{1}{\sqrt{2}}e^{x-c} - \frac{1}{\sqrt{2}}e^{-x-c} \,.
\label{phi6super1}
\ee
Each term separately would give no force, so it is the cross
terms that produce a non-zero result. Substituting into 
(\ref{tailsforce}), we find the repulsive force
\be
F = e^{-2c} \,,
\ee
independent of $X$. As the mass of the kink is $\quart$,
its acceleration is $4e^{-2c}$, and the mirror kink has the opposite
acceleration. Because the kink tail has exponentially small energy, 
the kink mass to the right of $X$ is effectively constant even as 
$X$ varies, and therefore an $X$-independent force is what's needed 
to produce a definite acceleration.  

The effective equation of motion for the kink is 
\be
\quart {\ddot c} = e^{-2c} \,,
\label{cmotion}
\ee
implying conservation of energy
\be
\quart {\dot c}^2 + e^{-2c} = {\rm const.}
\ee
From this one determines that the closest approach is $c = \log 2 - \log v$ 
if the kink and mirror kink approach from infinity with small speed $v$. 

Note that if we had decided on a different notion of the kink 
location, say $A = c + \delta$ with $\delta$ of order $1$, then we would 
have derived the equation of motion
\be
\quart {\ddot A} = e^{2\delta}e^{-2A} \,.
\label{Amotion}
\ee 
This shows that the coefficient in front of an exponentially small
force is only meaningful if one is careful to specify where the 
kink is located. Equations (\ref{Amotion}) and (\ref{cmotion}) are 
completely equivalent, although they seem to predict a closest
approach for the kinks differing by $\delta$.  

For a kink and antikink, one simply reverses the sign of the second term 
on the right hand side of (\ref{phi6super1}) and finds an attractive force
of the same strength.


\vspace{5mm}

\begin{thebibliography}{99}

\bibitem{Raj} R. Rajaraman, \textit{Solitons and Instantons},
Amsterdam, Elsevier Science, 1982.

\bibitem{book} N. Manton and P. Sutcliffe,
\textit{Topological Solitons},
Cambridge University Press, 2004.

\bibitem{Shn} Y.~M. Shnir,
\textit{Topological and Non-Topological Solitons in Scalar Field Theories},
Cambridge University Press, 2018.

\bibitem{BM} I.~V. Barashenkov and V.~G. Makhankov, 
Soliton-like ``bubbles'' in a system of interacting bosons,
\textit{Phys. Lett.} \textbf{A128}, 52 (1998).

\bibitem{Bo} E.~B. Bogomolny,
The stability of classical solutions, 
\textit{Sov. J. Nucl. Phys.} \textbf{24}, 449 (1976).

\bibitem{Loh} M.~A. Lohe,
Soliton structures in $P(\varphi)_2$,
\textit{Phys. Rev.} \textbf{D20}, 3120 (1979).

\bibitem{KCS} A. Khare, I.~C. Christov and A. Saxena,
Successive phase transitions and kink solutions in $\phi^8$, $\phi^{10}$
and $\phi^{12}$ field theories,
\textit{Phys. Rev.} \textbf{E90}, 023208 (2014).

\bibitem{BMM} D. Bazeia, R. Menezes and D.~C. Moreira,
Analytical study of kinklike structures with polynomial tails,
\textit{J. Phys. Commun.} \textbf{2}, 055019 (2018).

\bibitem{Ma5} N.~S. Manton,
An effective Lagrangian for solitons,
\textit{Nucl. Phys.} \textbf{B150}, 397 (1979).

\bibitem{Man} N.~S. Manton,
The force between 't Hooft-Polyakov monopoles,
\textit{Nucl. Phys.} \textbf{B126}, 525 (1977).

\bibitem{GE} J.~A. Gonz\'alez and J. Estrada-Sarlabous,
Kinks in systems with degenerate critical points,
\textit{Phys. Lett.} \textbf{A140}, 189 (1989).

\bibitem{MGGL} B.~A. Mello et al.,
Topological defects with long-range interactions,
\textit{Phys. Lett.} \textbf{A244}, 277 (1998).

\bibitem{BG} E. Belendryasova and V.~A. Gani,
Scattering of the $\varphi^8$ kinks with power-law asymptotics,
\textit{Commun. Nonlinear Sci. Numer. Simulat.} \textbf{67}, 414 (2019).

\bibitem{Chr} I.~C. Christov et al.,
Long-range interactions of kinks,
arXiv:1810.03590 (2018).

\bibitem{GR} I.~S. Gradshteyn and I.~M. Ryzhik,
\textit{Tables of Integrals, Series, and Products, 5th ed.},
ed. A. Jeffrey, Boston, Academic Press, 1994.

\bibitem{WWW} https://www.integral-calculator.com

\bibitem{Chr2} I.~C. Christov et al.,
Kink-kink and kink-antikink interactions with long-range tails,
arXiv:1811.07872 (2018).


\end{thebibliography}
\end{document}